\documentclass{aa}  

\usepackage{graphicx}
\usepackage{multirow}

\usepackage{amsmath}
\usepackage{natbib}
\usepackage[]{hyperref}
\usepackage{url}
\usepackage[usenames,dvipsnames]{color}  
\usepackage[title]{appendix}
\usepackage[normalem]{ulem}

\usepackage{txfonts}

\usepackage{caption}
\usepackage{subcaption}

\usepackage{bm}

\def\msun{{\rm ~M}_{\odot}}
\def\rsun{{\rm ~R}_{\odot}}

\def\mpy{{\rm ~M}_{\odot} {\rm ~yr}^{-1}}

\begin{document}

   \title{Testing the presence of a dormant black hole inside HR~6819}

   \author{A. Romagnolo
          \inst{1}
          A. Olejak
          \inst{1}
          A. Hypki
          \inst{1}
          G. Wiktorowicz
          K. Belczynski
          \inst{1}
          }

   \institute{
   Nicolaus Copernicus Astronomical Center, The Polish Academy of Sciences, ul. Bartycka 18, 00-716 Warsaw, Poland\\
              \email{amedeoromagnolo@gmail.com, chrisbelczynski@gmail.com}
             }

   \date{Received August 09, 2021; accepted August 22, 2022}

 
  \abstract
  {HR~6819 was recently reported to be a triple system with a non-accreting black hole (BH). The inner binary system was defined as a B3~III type star (a $5-7\msun$ star estimated to be at the end of its main sequence) and a dormant BH ($>4.2\msun$). The period of the inner binary was estimated to be $\sim 40$ days with an eccentricity in the range $0.02-0.04$. As the inner binary is not resolved, the third component may actually just be spatially coinciding with the inner binary.}
  {In this study we test whether the system's inner binary can be reconstructed using the isolated binary evolution in the Galactic field or through the dynamical evolution within globular star clusters. Our goal is to understand the formation of the HR~6819 inner binary.}
  {To simulate the inner binary evolution we assumed that the influence of the third body on the inner binary is negligible. We created synthetic populations of BH-main sequence binaries for the Galactic disc and the Galactic globular clusters to compare to the reported parameters of the HR~6819 inner binary. We have adopted very optimistic input physics, in terms of common envelope evolution and BH formation, for the formation of binaries similar to the reported inner HR~6819 binary.}  
  {Despite our optimistic assumptions we cannot form systems like the inner HR~6819 binary in globular clusters. Even with our extreme assumptions, the formation of an HR~6819-like binary in the Galactic field population is not expected.}
 {We argue that if a dormant BH actually exists in the reported configuration inside HR~6819, its presence cannot easily be explained by our models based on isolated and dynamical binary evolution.}
 

   \keywords{Binaries - Stars: black holes - Stars: individual: HR~6819}

\maketitle
%

\section{Introduction}
\label{sec:Intro}

Black holes (BHs) in binary systems are predicted to be around 10$\%$ of the whole Milky Way BH population, and most of them are BH-BH systems \citep{Olejak_2020}. Black holes with non-compact companions can be detected through interactions with them, but they are possibly a very small fraction of the total Galactic population. So far only $\sim$20 Galactic BH systems have been detected \citep{2007IAUS2383C,2014SSRv183223C}\footnote{\url{https://stellarcollapse.org/sites/default/files/table.pdf}}. Most of them are close X-ray binary systems transferring matter through an accretion disc. A recent review on stellar-origin BHs is provided in \citet{Bambi1806}. Black holes may also be detected as gravitational wave signal emitters \citep{Abbott1610} and in non-interacting binaries using astrometric and spectroscopic observations of their companion stars. Single BHs can potentially also be detected through microlensing \citep[e.g.][]{Wyrzykowski1605,2019ApJ8851W}, self-lensing \citep{2021arXiv210412666W} or while accreting from the interstellar medium \citep{Tsuna1806}. Theoretically Hawking radiation \citep{Hawking7403} may possibly allow for a detection of thermal radiation from the immediate vicinity of a BH.

According to recent simulations \citep{Sukhbold_2016,Patton_2020} stellar-origin BHs could either form from supernova events or from the direct collapse of the stellar progenitor without any emission of mass besides the neutrino emission. There is no hard line in terms of initial mass alone to whether an explosion or an implosion can happen, since this is rather dependent on factors such as the carbon-oxygen core mass and its initial uniform composition of carbon and oxygen, which in turn depend on the winds, mass loss, rotation, mixing effects, binary interactions and nuclear reaction rates during the early stages of stellar evolution (up to the core-helium burning phase) \citep{Patton_2020}. These simulations suggest that direct BH collapse is a very frequent phenomenon, more importantly for massive carbon-oxygen or iron cores and in general for stars between 10 and 30$\msun$ at the zero-age main sequence (ZAMS).

HR~6819 is reported to be a triple system harbouring a BH candidate \citep{Rivinius_2020}. The BH candidate mass was estimated on the basis of radial motions of the visible star in the inner binary. The spectral type of the companion star has been classified as B3~III. Using single-star evolutionary tracks \citep{Ekstrom1201} and observational data \citep{Hohle1004}, the most likely range for the star was estimated to be between 5 and $7 \msun$ and the star was estimated to be at the end of its main sequence (MS). The lower limit for the BH candidate mass was set at $4.2 \msun$. The star was reported to be roughly 1 $\msun$ more massive than its unseen companion, while the inner binary is supposed to have a period of $40.333\pm0.004$ and an eccentricity of $0.03\pm0.01$ . The outer orbit with the Be star (with a mass of $\sim6\msun$) is instead 
unconstrained. On top of that, the outer star does not show any significant RV variation and may actually be just spatially coinciding with the inner binary. Neither the Fiber-fed Extended Range Optical Spectrograph (FEROS) \citep{Rivinius_2020}, nor {\it Gaia} have a sufficient resolution to resolve the full triple system. The age of the whole triple system was estimated to be between 15 and 75 Myr. 

Other possible interpretations of the observational data cannot be excluded. For instance,
spectral features of HR~6819 may also be explained by a close binary in a 4-day period, 
composed of two A0 stars of $\sim$ 2.3 M$_\odot$ each with no BH~\citep{mazeh2020does}. 
In an independent analysis of the system spectrum, \citet{elbadry2021strippedcompanion} also propose the absence of a dormant BH. In their interpretation HR~6819 would be composed of two optical components: the B3 star and its Be stellar companion. 
Also \citet{refId02} propose a similar solution. According to their study HR~6819 could be a binary hosting a stripped B-type star and a rapidly rotating Be companion. Post-interaction lower-mass stars (exposed hot cores) can also mimic higher-mass isolated stars. The temperature of a B-type subdwarf can be higher than $20,000$ K despite having a mass of $\lesssim 1 \msun$ \citep[e.g.][]{Hever0909}. 

\cite{Safarzadeh_2020} also argue for the unlikelihood of a dormant BH inside the HR~6819 inner binary. Their work was based on three different considerations: (i) the small probability of having such a triple system configuration in terms of mass and spectral type in the Galactic population of stars, (ii) the extremely narrow range of orbital separations (both for the inner and the outer orbit) for the system to be dynamically stable, and (iii) the very narrow set of BH kicks and ejected mass at the BH formation that fits the observational constraints in terms of the orbital parameters for the inner binary. Additionally, they show that a chance alignment between a B III star and a Be star could be a potential explanation for the observational data.

\cite{stevance2021} pointed out, in their Letter regarding the potential existence of a dormant BH in NGC 1850, that the study of binary evolution in stellar populations can be beneficial for testing the strength of any claim regarding the presence of an unseen BH in binary systems. In our study we therefore try to reproduce the originally reported orbital parameters of HR~6819 within the framework of isolated binary evolution and dynamical formation in globular clusters (GCs). In both cases we test very optimistic models that can maximise the likelihood of reconstructing the reported parameters of the HR~6819 inner binary. In this way we will either set a best-case scenario (despite being unorthodox in terms of input physics) if we find any evolutionary solution for the reported parameters, or we will provide additional support for the absence of a dormant BH in HR~6819 in the case of no evolutionary solution. We assumed that HR~6819 is a triple system with the third component either unbound to the inner binary or with the outer orbit so wide that it does not affect its evolution. We consequently adopted the estimates from \citet{Rivinius_2020} as reference parameters of the HR~6819 inner binary. 

If we consider the influence of a third outer body, the inner binary might be gravitationally perturbed by the outer star during its evolution. Through the Kozai-Lidov mechanism \citep{Kozai_1962,Lidov_1962}, the two orbits can exchange orbital momentum (but not energy) and reciprocally alter their eccentricity and inclination. This effect could also produce highly eccentric inner binaries that could even merge before producing compact objects \citep{Naoz_2016}. Essentially, adding the outer star to the evolution of the system in our population study would result in a different eccentricity distribution as well as fewer BH-MS inner binaries arising due to the increased possibility of an early merger event.


\section{Method}
\label{sec:Method}

\subsection{\tt StarTrack}
\label{ST_M}
In our study we used the \textsc{StarTrack}\footnote{\url{https://startrackworks.camk.edu.pl}} population synthesis code \citep{Belczynski_2002,Belczynski_2008}, which allows the user to simulate the isolated binary evolution of the system with a wide variety of initial conditions and physical parameters. The most up-to-date description of \textsc{StarTrack}'s standard physical parameters and approaches, as well as the adopted model for star formation rates and the metallicity distribution of the Universe, can be found in \cite{2020A&A...636A.104B}, with two updates in Sect. 2 of \cite{OlejakFishabch2020}. 

We introduced a significant modification compared to the standard StarTrack assumptions, which is that all BHs (regardless of their mass) are formed via direct collapse without asymmetric mass ejection in supernova explosions. In such scenario, BHs do not receive natal kicks except for the ones due to asymmetric neutrino emission. The mass of the resulting compact object is equal to the mass of the pre-collapse star minus the emitted neutrino mass. This approach has also been adopted in \cite{Zapartas_2019}, where every single star beyond 20$\msun$ at ZAMS was assumed to not initiate a supernova event. We discriminate between whether the resulting compact object is a neutron star or a BH based on its final mass: if it is $\leq 2.5$ $\msun$, the compact object will be a neutron star, otherwise we will have a BH.

In our simulations we tested several scenarios with different values of metallicity Z (see Table~\ref{table:ParSp}), neutrino mass loss during BH collapse $F_{\rm loss}$ (in terms of percentage of total mass loss), efficiency of orbital energy loss for common envelope (CE) ejections, $\alpha$, the initial system semi-major axis, a$_{\rm{0}}$, and the stellar masses of progenitors at ZAMS, respectively M$_{\rm{ZAMS,a}}$ and M$_{\rm{ZAMS,b}}$. 

\begin{table}[ht]
\caption[Parameter Space]{Tested parameter space for the isolated binary evolution of HR~6819. }
\centering
\begin{tabular}{c c c}
\hline\hline 
Parameter & Tested Values\\ 
\hline
 Z & 0.01, 0.0105, 0.011, 0.015, 0.02 \\ 
 $F_{\rm loss}$ & 1\%, 6\%, 7\%, 10\%, 20\% \\
 $\alpha$ & 0.5, 1, 3, 4.5, 5 \\
 M$_{\rm ZAMS,a}$ [M$_\odot$]& 10,20,24,25,50,150 \\
 M$_{\rm ZAMS,b}$ [M$_\odot$]& 3, 6, 7, 8, 9, 10, 50, 150 \\
 a$_0$ [R$_\odot$]& 2000, 2500, 2650, 2655, 3000, 5000, 10000 \\
 \hline
\end{tabular}
\tablefoot{Here \textbf{Z}-metallicity, \bm{$F_{\rm loss}$}-neutrino mass loss during BH formation, \bm{$\alpha$}-efficiency of orbital energy loss for CE ejection, \bm{$M_{\rm ZAMS,a/b}$} - masses at ZAMS of respectively the BH progenitor star and its stellar companion, and \bm{$a_{\rm{0}}$} - initial semi-major axis.}
\label{table:ParSp}
\end{table}


\subsection{\tt MOCCA}
\label{MOC}

To investigate a possible influence of dynamical interactions on the formation of HR~6819 within Galactic globular star clusters we used the numerical simulations performed with the \textsc{mocca} code\footnote{\url{http://moccacode.net}}. It is currently one of the most advanced codes to simulate full stellar and dynamical evolution of real-size star clusters \citep{Giersz2014arXiv1411.7603G,Hypki2013MNRAS.429.1221H,Giersz1998MNRAS.298.1239G}. \textsc{mocca} is already used in a wide number of projects to investigate the evolution of compact binaries, for example BHs binaries \citep{Hong2020MNRAS.498.4287H}. Moreover, it is able to closely follow N-body codes and provides almost as much information about stars and binary 
stars \citep{Wang2016MNRAS.458.1450W}. The strong dynamical interactions are performed with the \textsc{fewbody} code \citep{Fregeau2004-01-004}.

In order to increase the reliability of the results from the possible dynamical channel, the \textsc{mocca} code was fully integrated with the \textsc{startrack} code. Now, \textsc{mocca} follows all the stellar and binary evolution processes with \textsc{startrack} as was done with \textsc{sse/bse}\footnote{\url{https://astronomy.swin.edu.au/~jhurley/bsedload.html}}\citep{Hurley2000MNRAS.315..543H,Hurley2002MNRAS.329..897H,Hurley2013} codes in previous \textsc{mocca} studies. In total $120$ numerical simulations of GCs were performed. In each simulation a cluster began its evolution with $6\times10^5$ initial objects (binaries and single stars), a $95\%$ binary fraction was assumed, and a metallicity of $Z=0.0105$ was adopted, with a $60$~pc tidal radius and a $2$~pc half-mass radius. Every simulation was started with a unique 'seed' value to generate different realisations of the initial conditions. The reason behind choosing one mass and one concentration was just simplicity. The goal was to check if the binary resembling HR 6819 can be formed with the help of dynamical interactions, which are characteristic of GCs (collisional systems). For that we decided to use a relatively massive initial cluster, with high density, in order to provoke various dynamical interactions. If the dynamical interactions could actually create many HR-like binaries, we would need to carefully select a broader range of initial conditions in order to find out in which GCs such systems are being preferentially formed. On the contrary, if dynamical interactions were shown to not have a role in the formation of a binary resembling HR~6819, there would not be the need to broaden the range of initial conditions.

All GC models were evolved with the same input stellar and binary physics (see Sect.~\ref{sec:MOCCA}) that increases possibility of HR~6819 inner binary formation (see Sect.~\ref{sec:OPS}). However, each model was a different realisation of the initial parameters for stars and binaries within their adopted initial distributions (star masses, binary orbits, initial positions, and velocities within a given cluster). This was done to ensure that we will not miss any potential HR~6819 formation channel due to poor statistics.


\section{Isolated binary evolution}
\label{sec:Results}

In order to reconstruct the HR~6819 inner binary, we had to recover some scenarios that, similarly to \cite{Rivinius_2020}, gave a BH-MS binary with component masses $>4.2\msun$ for the BH and $5-7\msun$ for the B star (the BH inner binary orbit companion). Additionally the two objects are supposed to differ in mass by $\sim 1\msun$ (the BH being less massive). Finally, the observational constraints require an orbital period of $40.3$ days and an eccentricity of $\sim 0.03$. From now on we will refer to the BH progenitor as the primary star and its B companion as the secondary star.

\subsection{Simulation}
\label{sec:OPS}

In isolated binary evolution calculations we can make an educated guess as to how we need to push evolutionary parameters to maximise the chance of formation of a binary similar to that of the inner binary of HR~6819. Our guess was based on several hundred trials of running evolution of massive binaries. We find that BH formation, CE evolution and the metallicity of a binary are important factors in the formation of BH-MS binaries with parameters resembling the reported architecture of HR~6819. We allow for some extreme changes in several parameters not excluded on theoretical grounds. Metallicity is allowed to change in the range $Z=0.01-0.02$ allowed for the Galactic disc, neutrino mass loss during BH formation is allowed to change in the range $F_{\rm loss}=1-20\%$, and the efficiency of conversion of orbital energy into the ejection of the stellar envelope during the CE evolution in the range $\alpha=0.5-5$. 
We tested the combination for all the initial conditions described in Table \ref{table:ParSp} to see if any of the retrieved BH-MS fits is similar to the reported HR~6819 configuration. To do so, we compute the Euclidean distance in the multi-dimesional space $M_{BH}$, $M_{MS}$, $e$ and $per$ by taking the HR~6819 observational constraints into account (BH mass $4.2-6\msun$, B star companion $5-7\msun$, orbital period $30-50$ days\footnote{We set an arbitrary confidence range of $10$ days around the observed value, despite this not corresponding to the actual margin of error ($0.004$ days) reported by \citet{Rivinius_2020}.}, eccentricity $0.02- 0.04$). This distance was calculated as the modulus of the normalised difference vector [$\Delta M_{BH}$, $\Delta M_{comp}$, $\Delta per$, $\Delta e$] between the simulated BH-MS binary and the reported HR~6819 parameters. The components of the aforementioned vector are equal to 0 if they are within their respective observational ranges. Otherwise, we took the absolute difference between the simulated parameter and the closest extreme in the observed range and divided it by the range of the observed quantity (1.8 $\msun$ for the BH mass, 2 $\msun$ for the companion mass, 20 days for the orbital period and 0.02 for the orbital eccentricity). For example, if $M_{BH}$ were 5 $\msun$, then $\Delta M_{BH}$ would be 0 because $M_{BH}$ is between 4.2 and $6\msun$; if it were instead 3 $\msun$, it would be closer to 4.2 $\msun$ (the observed minimum) rather than 6 $\msun$ (the observed maximum) and therefore $\Delta M_{BH}$ = |3 - 4.2|/1.8 $\sim$ 0.7 , while if it were 13 $\msun$, then $\Delta M_{BH}$ = |13 - 6|/1.8 $\sim$ 3.9 . In the following equations this procedure is explicitly shown:

\begin{equation}
\Delta M_{comp}= 
\begin{cases}
    0 & \text{if } 5 < M_{\rm comp}< 7\\
    min[|M_{\rm comp}-5|; |M_{\rm comp}-7|]/2,& \text{otherwise}\\
\end{cases}
\end{equation}

\begin{equation}
\Delta M_{BH}= 
\begin{cases}
    0 & \text{if } 4.2 < M_{\rm BH}< 6\\
    min[|M_{\rm BH}-4.2|; |M_{\rm BH}-6|]/1.8,& \text{otherwise}\\
\end{cases}
\end{equation}

\begin{equation}
\Delta per= 
\begin{cases}
    0 & \text{if } 30 < per< 50\\
    min[|per-30|; |per-50|]/20,& \text{otherwise}\\
\end{cases}
\end{equation}

\begin{equation}
\Delta e= 
\begin{cases}
    0 & \text{if } 0.02 < e< 0.04\\
    min[|e-0.02|; |e-0.04|]/0.02,& \text{otherwise}\\
\end{cases}
\end{equation}

The Euclidean distance from HR~6819 for a specific system will therefore be

\begin{equation}
    D = \sqrt{\Delta M_{BH}^2 + \Delta M_{comp}^2 + \Delta per^2 + \Delta e^2}
\end{equation}

Out of the 336000 tested binaries with different initial conditions, the Euclidean distance to HR~6819 of the ones that form a BH-MS binary at any point in their life and that have at least one of the four parameters within \cite{Rivinius_2020} observational ranges are plotted in Fig. \ref{fig:param_SP}. This difference can go up to $1.2\times 10^6$. For the sake of an understandable visualisation only the ones within a Euclidean difference of 15 are shown in the histogram.

\begin{figure}[!htbp]
\begin{subfigure}{.5\textwidth}\ContinuedFloat
\centering
\includegraphics[width=0.88\linewidth]{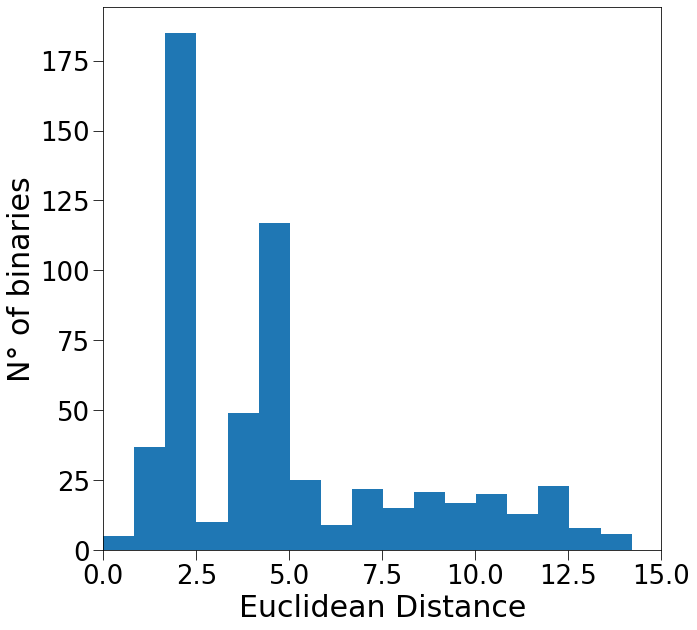}
\end{subfigure}
\caption{Histogram of the Euclidean distance distribution between the test binaries and the HR~6819 observational constraints with D < 15. Only two binaries have D = 0.}   
\label{fig:param_SP}
\end{figure}

Among them, only two have D = 0. They both have ZAMS masses of 24 $\msun$ (primary) and 7 $\msun$, $Z=0.0105$, $\alpha=5$ (as already used in other population studies, e.g. \cite{Santoliquido_2020}). One has $F_{\rm loss}=7\%$, and the other $F_{\rm loss}=6\%$.

Such a high CE efficiency does not imply energy that conservation has been violated, but rather that external energy sources that are not implemented in {\tt StarTrack} contribute to the energy balance. 
This can be explained as a contribution of the recombination energy of the envelope, which was shown to account for a fraction of the binding energy \citep{Kruckow2016,Fragos_2019}. It can also be due to a different prescription for the core--envelope boundary of the star just prior to the CE phase. The CE $\lambda$ prescription in {\tt StarTrack} comes from the so-called Nanjing $\lambda$ procedure \citep{Dominik_2012}, which comes from the \cite{Xu_2010} models. For evolutionary stages beyond the MS, these models are based on the assumption that the core--envelope boundary is located deep down in the star at the point where the hydrogen mass fraction, $X_H$, falls below 0.15 (the fits from \cite{Hurley2000MNRAS.315..543H}  are instead based on the detailed simulations from \cite{Pols_1998}, where the boundary is set at $X_H$ = 0.1). Instead, if the envelope were in reality on a shallower position, as suggested by \cite{Fragos_2019} and \cite{Marchant_2021} ($X_H$ = 0.3), it would mean that the {\tt StarTrack} $\lambda$ factor in the CE prescription has been underestimated. 
If the $\lambda$ prescription is kept fixed from the Nanjing $\lambda$ models, this underestimation can be replicated by setting $\alpha$ > 1.

We chose the initial setup with $F_{\rm loss}=7\%$ for our population study. We evolved $N=10^7$ massive binaries, in which the primary mass (the initially more massive binary 
component) is taken from the initial mass function three-part broken power law within a mass range of $18-150\msun$. 
The secondary component mass was instead taken from a flat distribution of mass ratio (secondary to primary) within the mass range $0.08-150\msun$. The orbital period was taken from \cite{Mink_2015}. We then scaled the total mass from our simulations to account for the whole $0.08-150\msun$ mass range for the primary stars by extending the initial mass function power law to primary stars between 0.08 and 18 $\msun$ at ZAMS (the secondary star mass is still taken from the flat distribution of mass ratio).

We can now select BH-MS binaries that resemble the reported inner binary in HR~6819. For this purpose we adopted the same Euclidean distance method that was described in Sect. \ref{sec:OPS}. In our simulations we find no binary (D = 0) resembling the reported parameters of the HR~6819 inner binary.  

\subsection{Calibration to the Milky Way} 
\label{sec.MWcal}

In our simulations we retrieved a number of BH-MS binaries equal to $2.4\times10^6$ and a total simulated ZAMS mass of $5.9\times10^6\msun$.
We then extrapolated our results to correspond to the entire Milky Way disc (isolated binary evolution). The disc mass was estimated by \cite{Licquia2015} to be 5.17 $\pm$ 1.11 $\times$ 10$^{10}M_\odot$. We therefore adopted a star formation rate of $\sim 5\msun$ yr$^{-1}$ and assumed that it has been constant in the disc over the last $10$ Gyr.
We needed to scale up (multiply) our results with the factor

\begin{equation}
f_{\rm scale,bi} = \frac{M_{\rm disc}}{M_{\rm sim}}  
\end{equation}

where $M_{\rm disc}=5\times10^{10}\msun$ is the total stellar mass of the Milky Way disc and, after being scaled to take into account the whole $0.08-150\msun$ mass range for primary stars, $M_{\rm sim}\sim4.2\times10^{9}\msun$ is the total simulated stellar mass in the case of 100\% binarity, and $M_{\rm sim}\sim6.9\times10^{9}\msun$ in the case of 50\% binarity, (i.e. two out of three stars in binary systems).

\begin{figure*}[!htbp]
    \centering
    \includegraphics[width=0.88\linewidth]{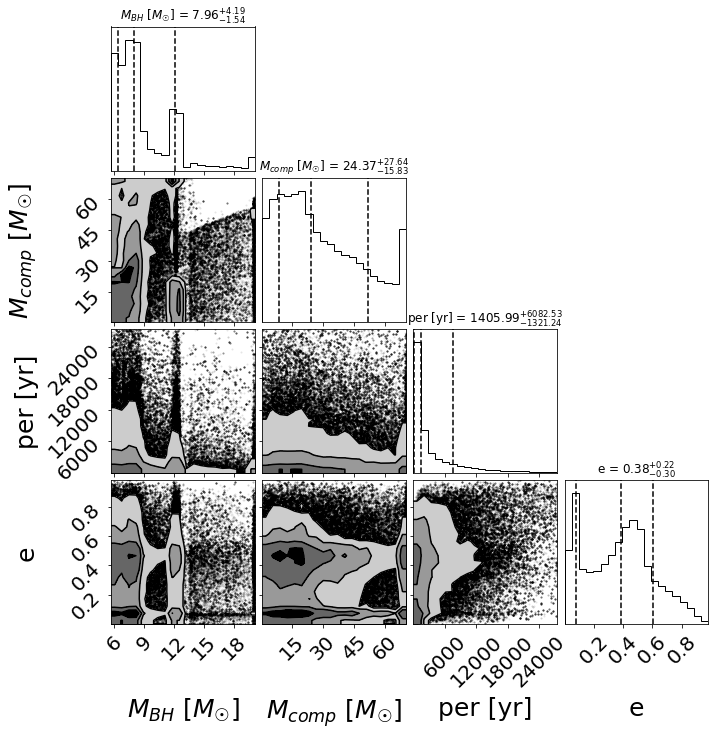}
    \caption{Corner plot of the correlations and predicted distributions of Galactic disc BH-MS binaries at 100\% binarity in terms of BH mass, MS companion mass, orbital period (within $\sim 28\times10^3$ years for better visualisation) and orbital eccentricity. The vertical dashed lines in the histograms show the 0.16, 0.5 and 0.84 quantiles for each parameter.}
    \label{fig:STCorner}
\end{figure*}

The above calibration results in a total number of BH-MS binaries that were formed in the Galactic disc over $10$ Gyr: $2.9\times10^7$ for 100\% binarity and $1.8\times10^7$ for 50\% binarity. Now we can distribute all BH-MS binaries uniformly over $10$ Gyr of disc evolution, and, knowing (from population synthesis calculation) each system BH-MS lifetime, we can estimate the current number of BH-MS binaries present in the Galactic disc: $427464$ for 100\% binarity and $249134$ for 50\% binarity. Neither of the two binarity cases shows any BH-MS system matching the reported HR~6819 configuration, which is expected since none of the evolved BH-MS binaries in our simulated population fit the observational constraints in the first place. As a reference for potential studies to test the reliability of our results, we also show in Fig.~\ref{fig:STCorner} a corner plot that includes the correlations between BH mass, MS companion mass, orbital period (within $\sim 28\times10^3$ years for better visualisation) and orbital eccentricity for the generated binaries at 100\% binarity, along with their distribution. Fig.~\ref{fig:chi} illustrates the current Galactic population of BH-MS binaries for 100\% binarity and their Euclidean distances from the reported HR~6819 binary parameters. For the sake of easy visualisation only the binaries at D<15 are shown in the histogram.

\begin{figure}[!htbp]
\begin{subfigure}{.5\textwidth}\ContinuedFloat
\centering
\includegraphics[width=0.88\linewidth]{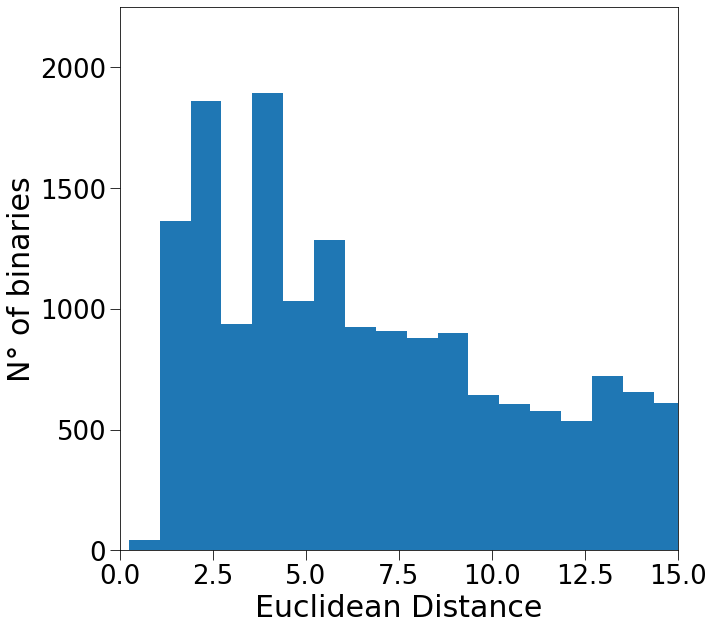}
\end{subfigure}
\caption{Current predicted population of Galactic disc BH-MS binaries and their Euclidean distance (in terms of orbital parameters) for 100\% binarity from the reported parameters of the HR~6819 inner binary \protect\citep{Rivinius_2020}.
No BH-MS binary out of $427464$ binaries resembles the HR~6819 inner binary to the point of having a Euclidean distance = 0.}   
\label{fig:chi}
\end{figure}

\subsection{Statistical significance of the formation of the HR~6819 inner binary}

Although we can find an evolutionary solution for the formation of the reported HR~6819 inner binary here we show that it is highly unlikely that such a binary would be expected in the current population of observed Galactic BH binaries. As pointed out in Sect.~\ref{sec.MWcal} we predicted up to $427464 \approx 4\times10^{5}$ BH-MS binaries currently present in the Galactic disc, with none resembling the reported HR~6819 inner binary.
Additionally, our estimate is done for very optimistic evolutionary assumptions on isolated binary evolution. We conclude that finding a binary similar to the reported inner HR~6819 binary in the current sample of Galactic BH systems is not expected.

Despite the fact that we do not consider it likely that such a system would form in the Galactic disc, in Appendix~\ref{sec.evol} we describe for completeness how such a system could
have possibly formed and evolved in the framework of the isolated binary evolution.
 

\section{Globular cluster evolution}
\label{sec:MOCCA}

After performing all $120$ GC simulations we find the formation of $89817$ BH-MS binaries. We find no BH-MS binary resembling the reported inner HR~6819 binary. The best candidate has a BH mass of $4.5\msun$, a companion MS star with mass of $6.7\msun$, an orbital period of $27.5$ days ($a=120\rsun$), and an eccentricity of $e=0.014$. The GC estimate ends here. We conclude that according to our models Galactic GCs do not form binaries resembling the reported inner binary of HR~6819. However, for consistency with the binary evolution estimate, we proceeded with an estimate of the current GC population of BH-MS binaries. 

We calibrated and extrapolated our results to represent the entire population of Milky Way clusters. The current total mass of the Milky Way GCs is based on GC mass estimates from \citet{Baumgardt2017MNRAS.464.2174B} \footnote{\url{https://people.smp.uq.edu.au/HolgerBaumgardt/globular/}} and equal to $3.7\times 10^7\msun$.  Mass loss of the star clusters through the Hubble time depends on many factors (e.g.  \citealt{Meiron2021MNRAS.503.3000M}). For the sake of simplicity, we assumed that half of the star mass of clusters was lost in their evolution. Thus, we arrive at an initial mass for Milky Way GCs of $7.4\times 10^7\msun$. On the other hand the stellar mass included in our $120$ GC simulation was $8.1\times 10^7\msun$. Therefore the multiplication factor to our results is 

\begin{equation} 
f_{\rm scale,gc} = \frac{7.4\times 10^7\msun}{8.1\times 10^7\msun}=0.9
\end{equation}

This gives us $8.2\times10^4$ BH-MS binaries formed by GCs in the Milky Way. However, only a fraction of them will survive until the present time, as GCs are old and most BH-MS stars that formed long ago are already gone or are no longer BH-MS binaries. 

\begin{figure*}
    \centering
    \includegraphics[width=0.88\linewidth]{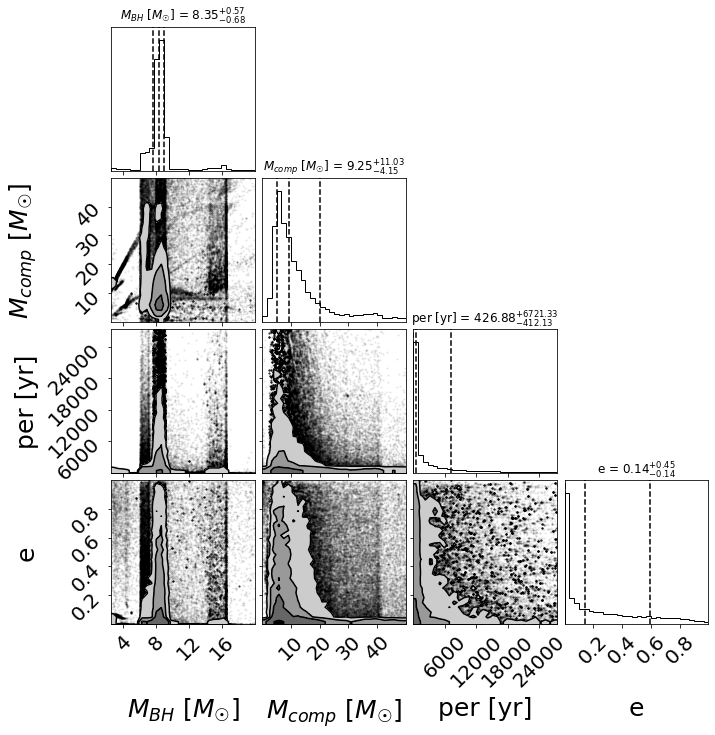}
    \caption{Corner plot of the correlations and predicted distributions of Galactic GC BH-MS binaries in terms of BH mass, MS companion mass, orbital period, and orbital eccentricity. The vertical dashed lines in the histograms show the 0.16, 0.5, and 0.84 quantiles for each parameter.}
    \label{fig:MOCCACorner}
\end{figure*}

Star formation in the GCs was taking place at large redshifts, $z>2$~\citep{Katz2013}. We, conservatively (in the context of survival of BH-MS binaries with rather massive MS stars), assumed that all stars in Galactic GCs formed at $z=2$, so $t\sim10$ Gyr ago. Since we know BH-MS lifetimes from our GC simulations, we can remove BH-MS binaries that do not survive to the present time. The corner plot in Fig.~\ref{fig:MOCCACorner} shows the correlation between the current BH mass (within $20 \msun$ for better visualisation), MS companion mass, orbital period (within $\sim 28\times10^3$ years) and orbital eccentricity in Galactic GCs. On the contrary, in the case of field binaries, the BH masses, the MS companion masses, and the orbital eccentricity are mainly localised within specific margins

\begin{figure}[!htbp]
\centering
    \includegraphics[width=1\linewidth]{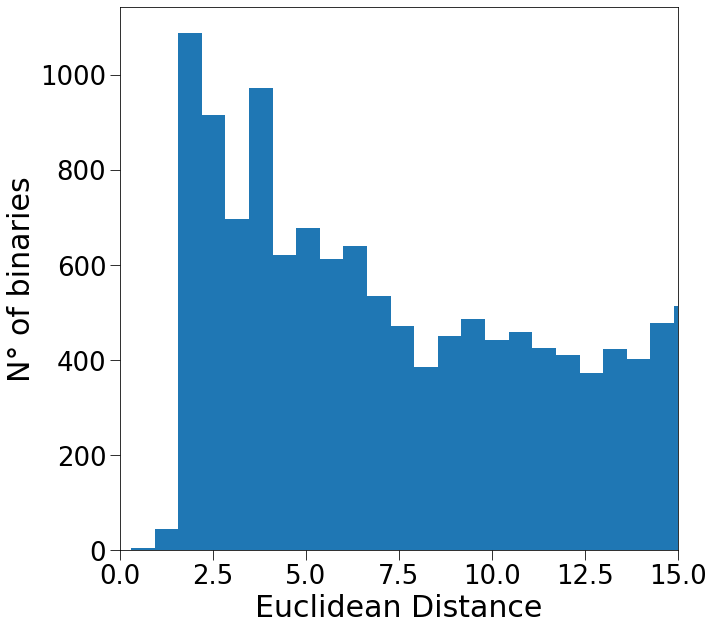}
     \caption{Current predicted population of Galactic GC BH-MS binaries and their Euclidean difference (in terms of orbital parameters) from the reported parameters of the HR~6819 inner binary \protect\citep{Rivinius_2020}.}

\label{fig:GCsCandidates}
\end{figure}

In Fig.~\ref{fig:GCsCandidates} we show the current population of GC BH-MS binaries.
We see that there is no system resembling the reported HR~6819 inner binary ($D=0$) and that there are very few systems with low Euclidean distance in orbital parameters from HR~6819. The main reason why it is hard to find dynamical candidates for the HR~6819 system in GCs comes from the fact that possible candidates can be found only at the very beginning of the star cluster evolution. Stars as massive as the companion to a BH ($>5\msun$) only live for very short time ($\lesssim 150$ Myr) and cannot contribute to the current GC (old) stellar populations. Even if a GC produced an HR~6819-like system, such a binary would be long gone. Thus, dynamical origin for a potential HR~6819-like system is highly unlikely. 


\section{Selection effects}
\label{sec:selection}

We analysed if the HR~6819 inner binary, despite its extremely low likelihood of arising through both isolated and dynamical binary evolution, could be more easily detected than other BH-MS binaries and therefore somehow compensate for the low formation probability.

So far, 69 electromagnetic BH binaries have been detected (from the BlackCAT catalogue\footnote{\url{https://www.astro.puc.cl/BlackCAT/transients.php}}: \cite{BlackCAT}), which is not yet a large enough sample from which to start statistical considerations. The problem becomes even more severe if we consider the BH discovery method. The majority of BH binaries are discovered through X-ray observations. Only five (plus the recent system in NGC 1850; \citealt{saracino2021}) have been discovered via RV measurements of the optical companion, which is the same method by which the HR~6819 BH was hypothesised to be found. On top of that, we cannot even be sure that NGC 1850 hosts a dormant BH, since \cite{El-Badry_2022} suggest that NGC 1850-BH1 is likely a stripped star and not a compact object. The first obvious conclusion here is that observations are more biased towards X-ray binaries than systems with dormant BHs, which therefore reduces the likelihood of HR-like systems being detected. In Table \ref{table:RV} we show the RV semi-amplitudes K1 for all the (hypothesised or confirmed) observed dormant BH-MS binaries. This was done to check whether HR~6819 has a considerably higher K1 that could compensate for an extremely low formation probability for isolated and dynamical binary evolution.

\begin{table}[ht]
\caption[RV]{RV semi-amplitudes for the observed BH-MS binaries.}
\centering
\begin{tabular}{c c c c}
\hline\hline 
Binary & K1 [km/s] & Reference\\ 
\hline
 NGC 1850 & 140.0 & \cite{saracino2021}\\
 HR~6819 & 61.3 & \cite{Rivinius_2020}\\
 LB-1 & 52.8 & \cite{Liu_2019}\\
 AS 386 & 51.7 & \cite{Khokhlov_2018}\\
 J05215658+4359220 & 44.6 & \cite{Thompson_2019}\\
 MWC 656 & 32.0 & \cite{Casares_2014}\\
 \hline
\end{tabular}
\label{table:RV}
\end{table}

\begin{figure}[!htbp]
\centering
    \includegraphics[width=1\linewidth]{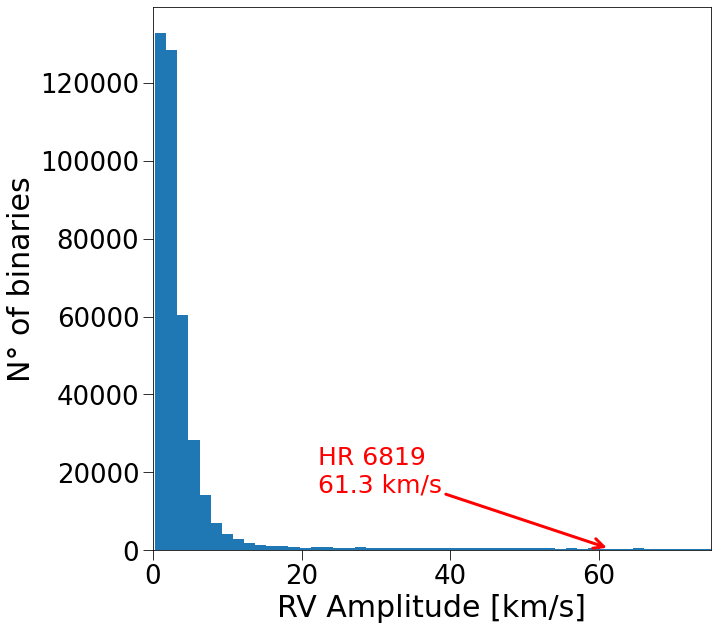}
     \caption{Current predicted RV semi-amplitudes for dormant BHs in the Galactic disc, with the observed HR~6819 K1 at 61.3 km/s. For the sake of easy visualisation the plot has been kept within semi-amplitudes of 80 km/s.}
\label{fig:RVAmpl}
\end{figure}

We also analysed the RV semi-amplitudes from our predicted current population of BH-MS binaries in the Galactic disc. Figure \ref{fig:RVAmpl} shows the K1 distribution for such binaries and the position of the HR~6819 inner binary.

All the observed dormant BH-MS binaries (be they confirmed or not) have a higher amplitude when compared with the average predicted distribution in the Galactic disc, which is understandable since they are therefore more likely to be detected with a high K1. To be more precise, 93\% of the binaries in our sample have a semi-amplitude below 60 km/s (roughly the semi-amplitude of HR~6819). Additionally, there was already admittedly an initial selection effect from the observations since HR~6819, among all the systems, was selected due to its line emission being similar to the one from LB-1 \citep{Rivinius_2020}, which is another system claimed to be a hierarchical triple system with a dormant BH in its inner binary \citep{Liu_2019,elbadry2021strippedcompanion}. This shows that there are indeed some selection effects that can contribute to increasing the detectability of HR-like systems, thought there are also some that would decrease it.

\section{Conclusion}

For the two major Galactic BH-MS formation channels, we adopted evolutionary physics that increases the possibility of formation of a system resembling the parameters of the HR~6819 inner binary reported by \citet{Rivinius_2020}. We tested several different types of initial input physics to retrieve the one that can maximise the possibility of an HR-like binary existing and therefore being detected. The evolution of stars in Galactic GCs does not allow for any avenues for creating such binaries. The Galactic disc also is a highly unlikely site for the formation of such binaries via isolated evolution.

We conclude that, even if a dormant BH existed in a triple system in a configuration similar to what has been proposed by \cite{Rivinius_2020}, its existence would be hard to explain with our models based on isolated and dynamical binary evolution. We nevertheless acknowledge that, despite the extremely low formation probability in our evolutionary channels, selection effects (both positive and negative) could alter the likelihood of such a system being detected.

\begin{acknowledgements}
AR, AO, AH and KB acknowledge support from the Polish National Science Center (NCN) grant Maestro (2018/30/A/ST9/00050). AH was also supported partially by Polish National Science Center grant 2016/23/B/ST9/02732. 
\end{acknowledgements}

\bibliographystyle{aa}
\bibliography{aamain}

\clearpage
\begin{appendices}
\counterwithin{figure}{section}                   

\section{Example of formation of HR~6819 inner binary in the framework of the isolated binary evolution}
\label{sec.evol}

\begin{figure}[ht]
    \centering
    \includegraphics[width=1\linewidth]{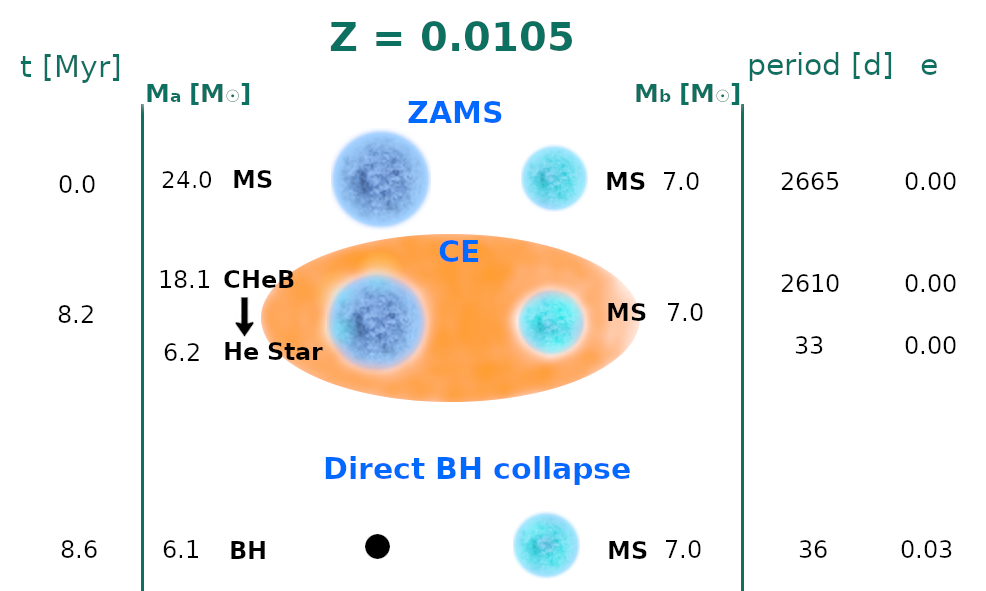}
    \caption[Evolution Diagram]{Formation of a system resembling the reported HR~6819 inner binary in the isolated binary evolution framework. It starts with two massive stars at metallicity $Z=0.0105$. While the companion star remains in its MS, the more massive star initiates a CE phase after $\sim 8.2$ Myr and shortly after it  collapses into a BH.}
    \label{Diagram}
\end{figure}

Here we present an example of one evolutionary scenario in the isolated binary evolution framework (Fig.~\ref{Diagram}) that we find leads to the formation of a binary that matches the observational constraints from \citet{Rivinius_2020}. Initially the system consists, at ZAMS, of two stars with masses of $24.0\msun$ and $7\msun$. All the input physics corresponds to our population synthesis.

\begin{figure}[ht]
    \centering
    \includegraphics[width=1\linewidth]{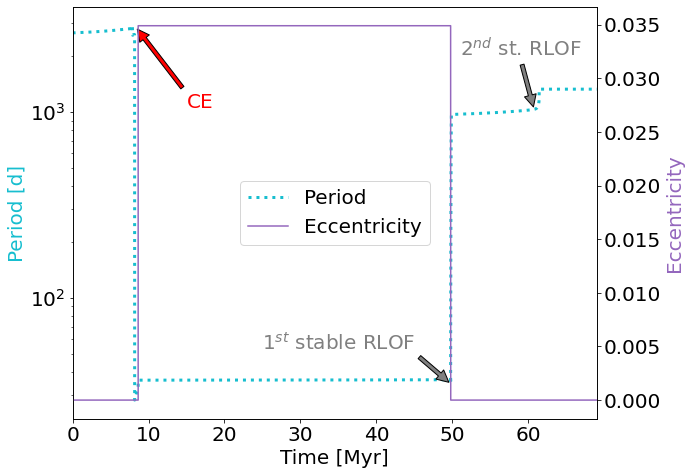}
    \caption[Eccentricity - Period]{Evolution of the binary period (blue) and eccentricity (purple) of the binary that forms a reported HR~6819 inner-binary-like system in the isolated binary evolution framework. The CE phase (red) takes place at $\sim 8.2$ Myr and was initiated by the primary star, which collapses shortly thereafter into a BH. A BH-MS system (similar to HR~6819) phase is noted for $\sim 40$ Myr ($\sim 10-50$ Myr after ZAMS). We also show the evolution of this binary beyond the BH-MS phase with two stable RLOF events (at $\sim 49$ Myr and $\sim 61$ Myr since ZAMS) initiated by an expanding secondary.}
    \label{EcPer}
\end{figure}

At the beginning the two stars are on a very wide, circular orbit (period $\sim 2.6 \times 10^3$ days, $e=0.0$). At $8$ Myr, the primary evolves off its MS, expands and evolves into a core-helium-burning giant. It initiates a CE phase, after which the orbital period is reduced to about $30$ days and the giant star loses its envelope. We then assume a direct collapse of the primary star into a BH. Due to the neutrino emission, the eccentricity of the system slightly increases to $e=0.035$. We note the formation of a BH-MS binary. The evolution of the orbital period and eccentricity of this binary is shown in Fig.~\ref{EcPer}. During the whole BH-MS phase (between $\sim 10-50$ Myr after system formation), the orbital period is $\sim 36$ d, which is close to the reported value of $40.3$ d. The eccentricity during the BH-MS phase, $e=0.035$, is in the observed [0.02; 0.04] range.

\begin{figure}[ht]
    \centering
    \includegraphics[width=1\linewidth]{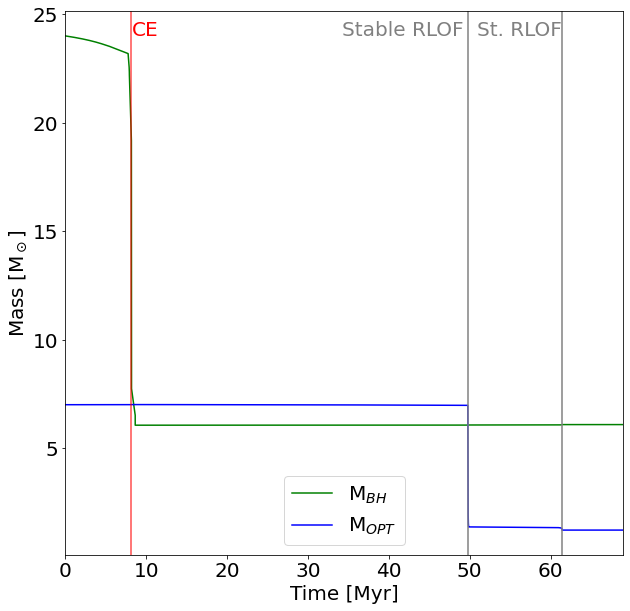}
    \caption[MassEvo]{Evolution of mass of the two components of the binary that forms reported HR~6819 inner-binary-like system in the isolated binary evolution framework. The primary (green line) loses most of its mass during the CE phase (red line) and collapses into a BH at $8.6$ Myr after ZAMS. The secondary star (blue line) remains in its MS with almost constant mass for $50$ Myr. Then it loses most of its mass during two stable RLOF events (grey lines).} 
    \label{MassEvo}
\end{figure}

The mass evolution of both components of this binary is shown in Fig.~\ref{MassEvo}. During the BH-MS phase, the BH mass is $6.1\msun$ and its MS companion's mass is $7.0\msun$. Both masses are within the range of observational estimates from \citet{Rivinius_2020}. During the BH-MS phase, the difference in mass between the BH and its companion is $\Delta M=0.9\msun$, which is consistent with the estimated value $\Delta M\sim 1\msun$ for HR~6819. 

\begin{figure}[ht]
    \centering
    \includegraphics[width=1\linewidth]{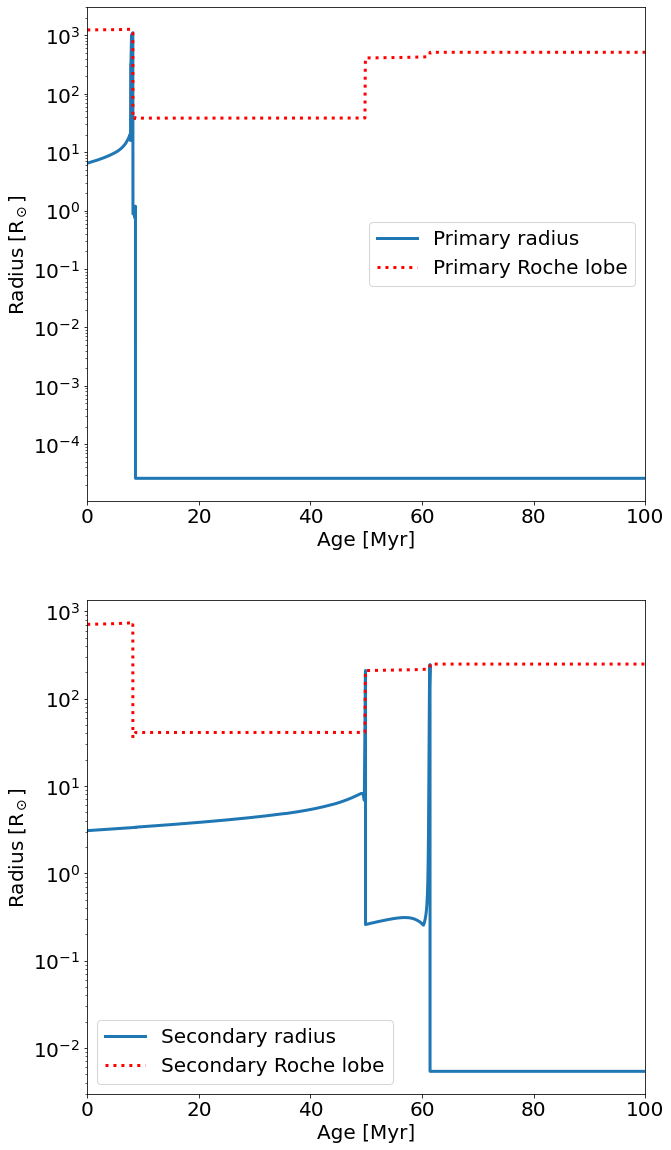}
    \caption[Radius]{Radial evolution of the two components of the binary that forms a reported HR~6819 inner-binary-like system in the isolated binary evolution framework. After the formation of a BH, the secondary is well within its Roche lobe during its entire MS lifetime ($<50$ Myr).} 
    \label{Radius}
\end{figure}

In Fig.~\ref{Radius} we show the temporal evolution of the radii of the two components for this binary in comparison with their Roche lobe radii. We note that during the BH-MS stage (the time between $8.6$ and $49.8$ Myr) there is no Roche lobe overflow (RLOF) event and that the wind mass loss from the secondary star ($\dot{M}<10^{-7}\mpy$) is too weak to produce any significant X-ray emission at the orbital separation found for this system (a $\sim108\rsun$) during this stage. The BH in our example is thus dormant and in agreement with the observational data.

In Figs. \ref{EcPer}, \ref{MassEvo}, and \ref{Radius} we also include predictions for the future evolution of this binary system. During the whole BH-MS phase, parameters of this binary remain almost unchanged, the only exception being the radius of the secondary star, which increases from $\sim 3\rsun$ to $\sim 6\rsun$ as it evolves through its MS.
The secondary leaves the MS at $\sim 50$ Myr and begins a rapid expansion that eventually leads to a RLOF phase. The system then enters a stable mass transfer phase that lasts for $\sim 0.1$ Myr. After the secondary loses its H-rich envelope (the BH does not accrete any significant amount of mass as accretion proceeds on a fast thermal timescale), it becomes a naked He star with a mass of $1.37\msun$. The RLOF and associated mass and angular momentum loss from the binary leads to significant orbital expansion to an orbital period of $\sim 10^3$ days (a $\sim 2.3\times10^4\rsun$) and the orbit becomes circularised. The low-mass helium star then undergoes expansion during shell He-burning and initiates a second stable RLOF at $\sim 61$ Myr since ZAMS. The loss of angular momentum again expands the orbit to an orbital period of $1279$ days ($\sim 2.1\times10^5\rsun$).
By the end of both RLOF episodes the secondary star has lost more than 80$\%$ of its mass and has cooled to become a carbon-oxygen white dwarf. We note the formation of a wide (coalescence time much larger than Hubble time) BH-white dwarf binary.

\end{appendices}

\end{document}